\documentclass[a4paper]{jpconf}  
\usepackage{graphicx}
\usepackage{subfigure}
\usepackage{color}

\begin{document}

\title{Potential mechanical loss mechanisms in bulk materials for future gravitational wave detectors}

\author{D Heinert$^1$, A Grib$^{1,2}$, K Haughian$^3$, J Hough$^3$, S Kroker$^{4}$, P Murray$^3$, R Nawrodt$^{1,3}$, S Rowan$^3$, C Schwarz$^1$, P Seidel$^1$, A T\"unnermann$^4$}

\address{$^1$ Institut f\"ur Festk\"orperphysik, Friedrich-Schiller-Universit\"at, 07743 Jena, Germany}
\address{$^2$ Physics Department, Kharkiv V. N. Karazin National University, 61077 Kharkiv, Ukraine}
\address{$^3$ Institute for Gravitational Research, Department of Physics and Astronomy, Glasgow University, G12 8QQ Glasgow, UK}
\address{$^4$ Institut f\"ur Angewandte Physik, Albert-Einstein-Str. 15, 07745 Jena, Germany}

\ead{daniel.heinert@uni-jena.de}

\begin{abstract}

Low mechanical loss materials are needed to further decrease thermal noise in upcoming gravitational wave detectors. We present an analysis of the contribution of Akhieser and thermoelastic damping on the experimental results of resonant mechanical loss measurements. The combination of both processes allows the fit of the experimental data of quartz in the low temperature region (10\,K to 25\,K). A fully anisotropic numerical calculation over a wide temperature range (10\,K to 300\,K) reveals, that thermoelastic damping is not a dominant noise source in bulk silicon samples. The anisotropic numerical calculation is sucessfully applied to the estimate of thermoelastic noise of an advanced LIGO sized silicon test mass.

\end{abstract}

\section{Introduction}
The detection of gravitational waves demands a very high precision length measurement. Even strong gravitational wave sources will just cause an effect in the order of 10$^{-19}$\,m on earth. In modern interferometric gravitational wave detectors thermal noise significantly limits the sensitivity. Especially in the most sensitive frequency range from 10\,Hz - 200\,Hz thermally driven random vibrations of the surface of the optical substrates become important. Besides the classical Brownian noise \cite{Einstein1906,Gonzalez1994} other noise mechanisms have been identified to contribute to the total thermal noise: thermoelastic and photothermal noise \cite{braginsky1999} as well as thermo-refractive noise \cite{braginsky2000,Evans2008}.
All noise processes can equivalently be described by a corresponding process consuming energy. The link between noise and loss is given by the fluctuation-dissipation-theorem \cite{callen1951}.
Consequently knowing the dissipative properties, i.e. the temperature and frequency behaviour of all loss processes, of a system allows the computation of its thermal noise.

In the case of a gravitational wave detector the displacement noise of the surface is connected with the dissipation of mechanical energy within the material. Therefore, the mechanical loss of different substrate materials has been studied in the past, e.g. \cite{McGuigan1978,Crooks2004,Amico2002,Nawrodt2008,Penn2006}, to understand the origin of the mechanical dissipation. All these studies rely on ring down measurements of cylindrical bulk samples at their resonances. The estimated characteristic ring down time is finally used as a measure for the mechanical loss.

The results of these measurements consist of a superposition of many different internal loss processes. In order to understand or even minimise these different loss contributions it is necessary to separate them within the measured data sets.
Therefore a subsequent analysis aims at the identification of the different mechanical loss processes within the material. 
Fundamental intrinsic loss processes in bulk materials are thermoelastic damping and Akhieser damping, which is the result of phonon-phonon-interaction.
We will focus on a detailed description of these fundamental loss processes within this paper. 
Due to the small height of the cylindrical samples an isotropic plate theory \cite{bishop1997} is applied for evaluating the effect of thermoelastic damping.
Furthermore, a finite element analysis is developed to extend the theory of thermoelastic loss to anisotropic samples. This treatment allows an easy application to arbitrarily shaped test masses as well and helps identifying the loss sources in the ring-down experiments. For silicon we examplarily compare our approach to the isotropic plate model.

The second part of the paper applies the developed code to the calculation of thermoelastic noise of a gravitational wave detector bulk material having anisotropic material properties.

\section{Mechanical loss processes}
\subsection {Thermoelastic damping}

A test sample under vibration exhibits regions being compressed and regions being expanded during a vibrational cycle. Accordingly, the described volume change is linked to a temperature change by means of the coefficient of thermal expansion. The arising temperature gradient within the sample leads to a heat flow and, consequently, increases the entropy of the system \cite{braginsky1999} which leads to dissipation. This dissipation process causes the vibrational amplitude to decrease with time and is known as thermoelastic damping (TED). A non-vanishing coefficient of thermal expansion only occurs in a material if the interatomic potential has anharmonic contributions and thus TED is classified as a nonlinear effect of the lattice.

In order to calculate the TED numerically the mode shape of the resonant vibration of the substrate is required. 
With the knowledge of the mode shape one can estimate the volume change within the sample under deformation. This volume change is directly linked to the stress tensor $\sigma$. Quantitatively the temperature distribution $T$ within the sample is obtained by using the equation of heat transport (see e.g. \cite{Norris2005}):

\begin{equation}
C_p \dot{T} -\lambda_{ij}\frac{\partial T}{\partial x_i \partial x_j}=-\alpha_{ij}\dot{\sigma}_{ij}T_0 \ .
\label{equ:heat}
\end{equation}

In this equation $\lambda$ and $\alpha$ represent the tensors of thermal conductivity and thermal expansion respectively. Furthermore $C_p$ is the specific heat per unit volume and $T_0$ the average temperature of the sample.
In Eq.\,(\ref{equ:heat}) the Einstein convention is used, i.e. to sum over double indices.

Summing the work done by the new thermal field against the original stress field finally gives the dissipated energy $\Delta E$ per oscillation cycle \cite{Lifshitz2000,Fejer2003}. 
By using the complex notation for harmonic processes one obtains:
\begin{equation}
	\Delta E=\pi \int_V \alpha_{ij} \hat{\sigma}_{ij} \Im\left( \hat{T}\right) dV \ ,
\end{equation}
where $\Im(\hat{T})$ is the imaginary part of the temperature amplitude $\hat{T}$ ($T=\hat{T}e^{i\omega t}$) which is taken as the solution from Eq.\,(\ref{equ:heat}). This part of the temperature change is phase shifted compared to the mechanical stress.
The mechanical loss $\phi$ is then given by:
\begin{equation}
\phi=\frac{1}{2\pi} \frac{\Delta E}{E} \ 
\end{equation}
with the total vibrational energy $E$. 

The presented scheme allows the full anisotropic treatment of thermoelastic damping in solids.

\subsection{Akhieser damping}

The second important contribution to the mechanical damping in bulk samples results from the interaction of acoustic waves with thermal phonons.
Thermal phonons carry the thermal energy of the sample and obey a Planck distribution in thermal equilibrium. 
In 1939 Akhieser \cite{akhieser1939} developed a theory for this effect, where he treated the acoustic wave as a homogeneous stress background for the thermal phonons.
In the field of resonant loss measurements the homogeneous stress background is provided by the excited eigenmode of the sample.
This approximation requires that the lifetime of thermal phonons $\tau$ is short compared to the period $1/f$ of the acoustic vibration. 
For all materials and temperatures considered in this paper this condition ($f\tau \ll 1$) is well satisfied. 
At a frequency of 25\,kHz and at a temperature of 10\,K $f\tau$ reveals to be $6\times10^{-3}$ for silicon and $2.2\times10^{-3}$ for quartz.

Mechanical stress leads to a change of the phonon energies in the crystal via the Gr\"uneisenparameter $\gamma$.
In general each phonon branch shows a distinct parameter and therefore a distinct energy shift.
Consequently, the whole phonon ensemble is brought out of thermal equilibrium by an external stress.  
The thermodynamical equilibrium state is characterized by the maximum of entropy. 
Any process reestablishing equilibrium, here the collision between phonons, will, therefore, increase the entropy and, thus, dissipate energy.
The speed of phonon redistribution is limited by the phonon lifetime $\tau$ as the typical time between interactions.

Again Akhieser damping is a nonlinear effect, because without having anharmonic contributions in the interatomic potential phonon-phonon-interaction would be forbidden.

Akhieser losses $\phi_{ph}$ for the bulk vibrating solid are given by the following expression \cite{braginskybook}


\begin{equation}
	\phi_{ph}=\frac{C_p T \gamma^2}{\rho v_s^2} \frac{\omega \tau}{1+ \omega^2\tau^2}  \ .
	\label{equ:boemmel}
\end{equation}
In this model the Akhieser loss $\phi_{ph}$ depends on the sound velocity $v_s$, the angular freqency of the acoustic wave $\omega$ and the density $\rho$ of the material.

\section{Numerical estimation and application to experimental data}
\subsection{Crystalline quartz}

Akhieser and thermoelastic damping have been applied to the measured loss values of a crystalline quartz sample. 
The sample has a cylindrical shape (z-axis parallel to the cylinder axis) with a diameter of 76.2\,mm and a height of 12\,mm. 
The mechanical loss above 25\,K is dominated by processes which can be described in the model of double well potentials \cite{zimmer2007}. 
Former investigations identified interstitial aluminum together with alkali atoms to cause mechanical loss in quartz \cite{fraser1964} at temperatures above 25\,K. 
For the defect model does not satisfactorily explain the low temperature data, one has to take further mechanical loss processes into consideration.

In this paper we concentrate on the temperature region below 25\,K to exclude the influence of defects and to explain the behaviour with a combined model of TED and Akhieser damping. 
For a detailed description of this model see \cite{Gryb2009}.
Due to the small height of the sample compared to its diameter we used the approach of Bishop and Kinra \cite{bishop1997} to estimate TED in a plate. 
Mechanical losses due to Akhieser damping were calculated with the use of Eq.\,(\ref{equ:boemmel}) into which we inserted the material properties of quartz from the literature \cite{touloukian}. 
 
The theoretical values using both TED and Akhieser damping are compared to the measured loss values in Fig.\,\ref{fig:quartz} for an eigenmode at f\,=\,11.7\,kHz. 
It is seen that we obtained a good quantitative agreement between experiment and theory for the quite reasonable value of the Gr\"uneisen parameter $\gamma$\,=\,2.
This is an indication that the low temperature loss in quartz can exclusively be described by these two mechanisms.

At the experimental frequency of 11.7\,kHz the contribution of Akhieser losses is negligibly small (see Fig.\,\ref{fig:quartz}). However, when the eigenfrequencies exceed a few MHz, the contribution of Akhieser losses will become dominating.   

\subsection{Crystalline silicon}
\begin{figure}[tb]
\begin{minipage}[t]{18pc}
 \resizebox{!}{14pc}{\includegraphics{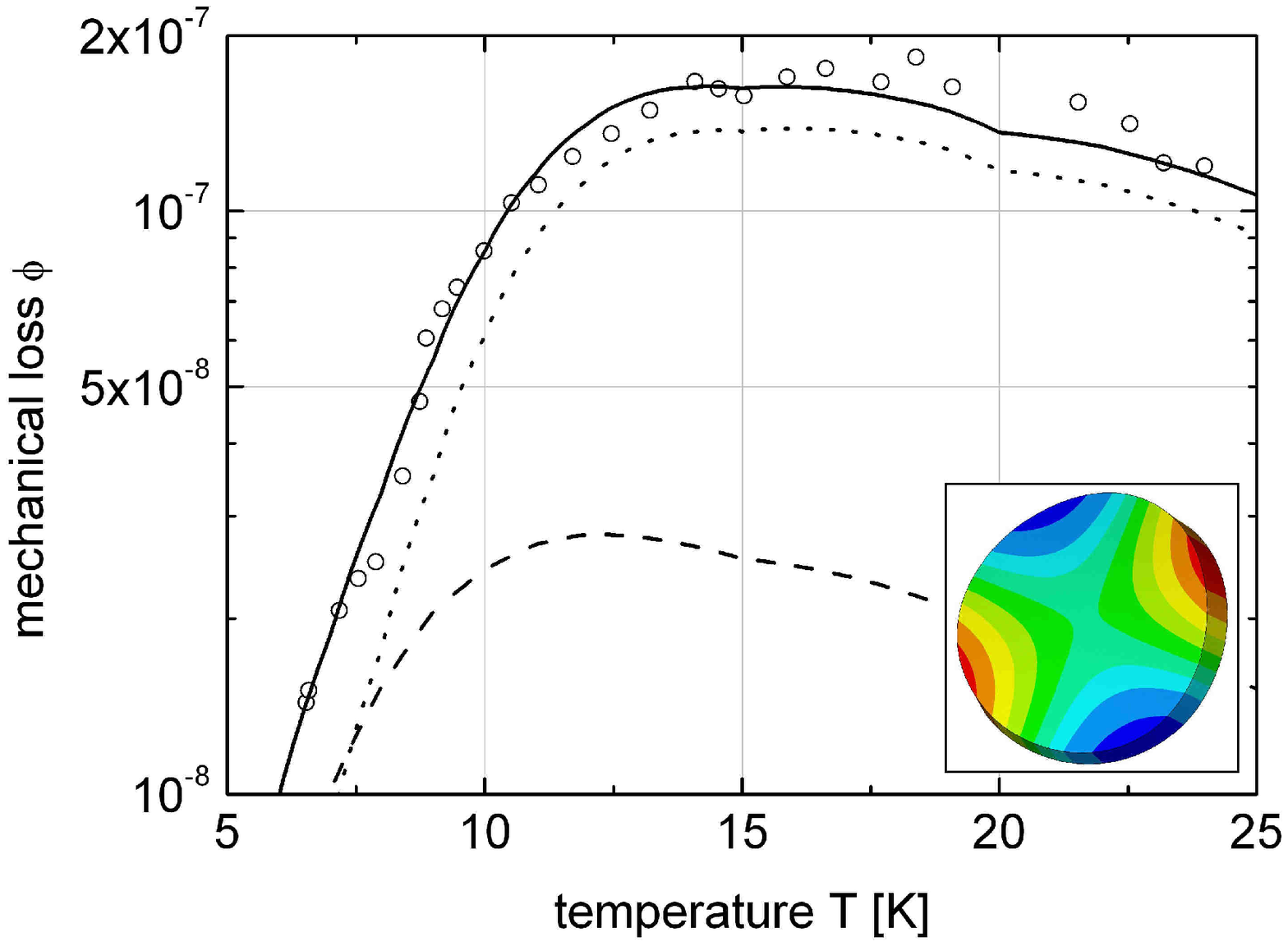}}
\caption{\label{fig:quartz}Measured mechanical loss (\opencircle) of a \mbox{\O~76.2\,mm} \mbox{$\times$~12\,mm} quartz sample in z-cut at f\,=\,11.7\,kHz. Calculated mechanical loss (\full) as the sum of TED (\dotted) and phonon damping (\dashed) as described in the text. In the lower right corner the deformation of the substrate in the cylinder axis is shown.}
\end{minipage}\hspace{2pc}%
\begin{minipage}[t]{18pc}
	\resizebox{!}{14pc}{\includegraphics{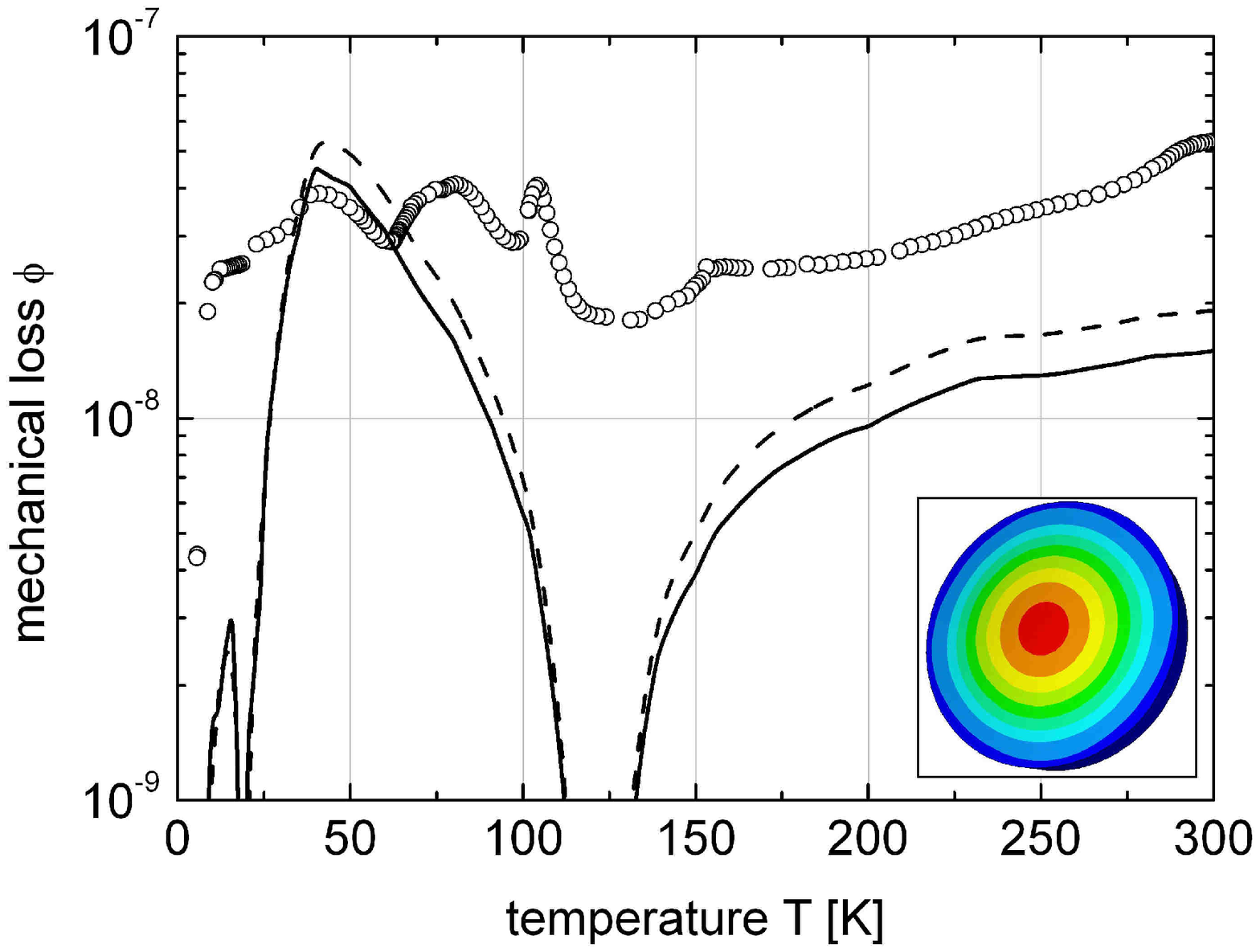}}
\caption{\label{fig:silicon}Measured mechanical loss (\opencircle) of a \mbox{\O~76.2\,mm} \mbox{$\times$~12\,mm} silicon sample in (100) direction at f\,=\,24.1\,kHz. The data are compared to TED obtained by a isotropic model (\full) and our anisotropic approach (\dashed). In the lower right corner the deformation of the substrate in the cylinder axis is shown.}
\end{minipage} 
\end{figure}

We also investigated silicon, the most promising candidate for future cryogenic gravitational wave detectors \cite{Winkler1994,Rowan2005,Nawrodt2008}. 
The cylindrical sample has the same geometry as the quartz sample (\mbox{\O~76.2\,mm} \mbox{$\times$~12\,mm}) and was oriented along the cylinder axis in (100) direction.
Again TED in silicon was at first calculated with the isotropic plate theory described in \cite{bishop1997}. 
We used the FE software COMSOL \cite{comsol} for a completely numerical and anisotropic calculation of TED. 
Consequently, the results of both models are compared in Fig.\,\ref{fig:silicon}. Thermophysical properties of silicon are taken from the references \cite{touloukian, hull}. 

For the given silicon resonance we encounter deviations between the anisotropic and isotropic model of up to 20\%. However, TED alone is not capable of explaining the measured loss data over the whole temperature range.
The results show, that TED significantly affects mechanical loss in our bulk-like sample only at a small temperature region around 40\,K. There the calculated dissipation even exceeds the experimental values. The origin of this discrepancy could be found in the deviation of the real sample parameters from the table values used in the calculation.


The Akhieser damping strongly depends on the thermal conductivity (see \cite{braginskybook}) which is in turn strongly dependent on the impurity concentration. Unfortunately, for the sample, used in this paper, this parameter is not known to a degree which would justify any estimate on Akhieser damping. Further investigation of the effect of phonon interaction on damping in silicon is on the way.

%

\section{Thermal noise due to TED in a gravitational wave detector}

\label{sec:noise}
The effect of TED was first outlined in 1937 by Zener for vibrating reeds \cite{zener1937}. 
In 1999 the importance of TED as a serious noise contribution in gravitational wave detectors was pointed out in \cite{braginsky1999}. This behaviour is especially relevant for crystalline test masses as sapphire and silicon which are proposed to be used in future detectors \cite{lcgt,et}. A refined model was developed in order to cover finite test mass sizes and non-adiabatic conditions \cite{cerdonio2001}. These conditions become important when small beams (e.g. in laser stabilization cavities), high thermal conductivity materials or low frequencies are used.

Using Levin's theorem \cite{levin1998} the resulting thermoelastic noise of the mirror is obtained by the calculation of the losses due to a virtual mechanical pressure applied  to the front surface of the sample. The spatial distribution of pressure along the surface $p(\vec{r})$ of the sample has to mimic the intensity distribution of the laser beam:

\begin{equation}
	p(\vec{r})=\frac{F_0}{r_0^2} \exp \left(- \frac{r^2}{r_0^2}\right) \ ,
\end{equation}
where $F_0$ is an arbitrarily chosen force amplitude of the excitation. The beam radius $r_0$ indicates the point of the 1/e decay of the laser intensity. 
 
\begin{figure}[htb]
\includegraphics[width=16pc]{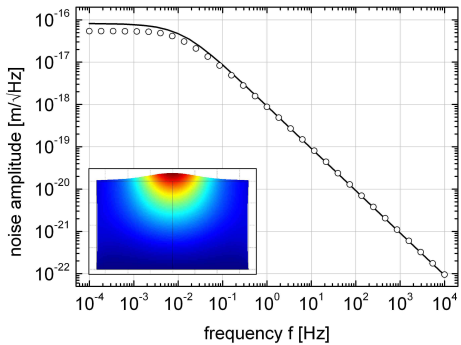}\hspace{1pc}%
\begin{minipage}[b]{20pc}\caption{\label{fig:cerdonio}Numerical estimate (\opencircle) of thermoelastic noise in an isotropically modelled silicon substrate with advLIGO geometry (\mbox{\O~340\,mm} \mbox{$\times$~220\,mm}) in comparison to the analytical treatment given in \cite{cerdonio2001} for the same substrate (\full). For both calculations a beam waist radius of $w_0=\sqrt{2}r_0=\mbox{60\,mm}$ and a temperature T\,=\,300\,K were used. The full anisotropic treatment leads to the same result as the isotropic one. Thus, the plot is omitted. The inset shows the deformation due to the Gaussian pressure profile.}
\end{minipage}
\end{figure}

The spectral noise power density $S_x^2(f)$ due to surface fluctuations then results from \cite{levin1998}:
\begin{equation}
	S_x^2(f)=\frac{2k_BT}{\pi^2 f^2}\frac{W_{diss}}{F_0^2} \ .
	\label{equ:noise}
\end{equation}
In Eq.\,(\ref{equ:noise}) $W_{diss}$ describes the energy dissipated per cycle due to all relevant loss mechanisms. In order to obtain thermoelastic noise the dissipation due to TED needs to be considered. Applying the pressure $p(\vec{r})$ in the quasistatic approximation allows the calculation of $W_{diss}$ with the FE software COMSOL. Fig.\,\ref{fig:cerdonio} compares the full anisotropic FEA with the results obtained by using the analytical calculation in \cite[Eq. (20)]{cerdonio2001}. 

At first we compared the thermoelastic noise level given by the analytical and our numerical theory for an isotropic body of finite size (advLIGO geometry). 
Both results exhibit a good agreement although the analytical calculation is strictly valid for infinite test masses.
This effect of finite sized geometry explains the encountered difference below 5\,\% over the frequency range of 1\,Hz to 10\,kHz between the models.  
An analytic expression for the effect of finite sizes \cite{liu2000} on isotropic test masses also explains changes in the order of 5\,\%.
The only major discrepancy occurs at low frequencies where the difference in the boundary conditions become important.
 
Our analysis showed that introducing anisotropy into the problem only leads to a minor change in the noise level of less than one per cent. Thus, in the quasistatic model far away from any resonances, for silicon the effect of anisotropy on thermoelastic noise is negligible. Close to the resonances this assumption breaks down and a full anisotropic treatment is needed. Although this is not necessary for the optical components (where the resonances are always well outside the detection band) this might become important for the suspension elements having resonances inside the operation frequency band.

\section{Conclusion}
Considering both B\"ommels model for phonon damping and the model of Bishop and Kinra for thermoelastic damping in plates allows the interpretation of the low temperature behaviour of mechanical losses in quartz. We showed a quantitative agreement with our measurement for the temperature interval from around 10\,K to 25\,K. In contrast, the repeated analysis for a bulk silicon sample revealed that thermoelastic damping is not to be the dominating factor within the experimental temperature interval from 5\,K to 300\,K. TED alone does not explain the measured data. Therefore we expect other loss mechanisms to contribute essentially to the mechanical loss within the considered temperature region. 

Finally the comparison of our noise calculation to an analytical approach proved the reliability of our numerical code. 
This code allows to include arbitrary sample geometries as well as the anisotropic character of the material.
Its application on bulk silicon shows, that the influence of the anisotropy on thermoelastic noise can be neglected for this material for typical mirror geometries. 

Nevertheless, in the field of mechanical loss measurements on samples of varying size the effect of anisotropy is important in order to seperate different loss mechanisms. Thus, the developed code is a valueable tool to include anisotropy into the mechanical loss data analysis as well as the thermal noise calculation for small structures.

\ack
This work was in part supported by the German science foundation DFG under contract SFB TR7.
\section*{References}
\bibliography{AMALDI8}

\end{document}